\documentclass{elsart3p} 

\usepackage{bm}

\newtheorem{theorem}{Theorem}
\newcommand{\bc}{{\bm c}}
\newcommand{\bx}{{\bm x}}
\newcommand{\bu}{{\bm u}}
\newcommand{\bM}{{\bm M}}
\newcommand{\expo}[1]{\exp\left[#1\right]}
\newcommand{\expoa}[1]{\exp\left(#1\right)}

\begin{document}

\begin{frontmatter}

\title{Fundamental Conditions for $N$-th Order Accurate Lattice
  Boltzmann Models}
 
\author{Hudong Chen and Xiaowen Shan}

\address{EXA Corporation, 3 Burlington Woods Drive, Burlington, MA
  01803, USA}

\begin{abstract}

In this paper, we theoretically prove a set of fundamental conditions
pertaining discrete velocity sets and corresponding weights. These
conditions provide sufficient conditions for {\em a priori}
formulation of lattice Boltzmann models that automatically admit
correct hydrodynamic moments up to any given $N$-th order.

\end{abstract}

\begin{keyword}
Lattice Boltzmann \sep hydrodynamic moments

\end{keyword}

\end{frontmatter}

\section{Introduction}

Lattice Boltzmann methods (LBM) has been recognized as an advantageous
numerical method for performing efficient computational fluid
dynamics~\cite{Benzi,ChenDoolen}. Not only it offers a new way of
describing macroscopic fluid physics, but also it has become a
practical computational tool and already has been making substantial
impact in real world engineering
applications~\cite{science}. Furthermore, according to a more recent
interpretation, LBM models are special discrete approximations to the
continuum Boltzmann kinetic equation~\cite{Shan98,Shan06}. Owing to
such an underlying kinetic theory origin, LBM is expected to contain a
wider range of fluid flow physics than the conventional hydrodynamic
fluid descriptions~\cite{cerci,gad,agar,macro}.  The latter, such as
the Euler or the Navier-Stokes equation, rely on various
``theoretical'' closure approximations for the non-equilibrium effects
that are problematic when deviations from local thermodynamic
equilibrium are no longer considered small. In addition, due to the
fact that the fundamental turbulence modeling is built upon an analogy
to regular fluid flows at finite Knudsen numbers, a kinetic theory
representation is argued to be more suitable than the classical
modeling approach of modified Navier-Stokes
equations~\cite{jfm}. However, how much the original range in kinetic
theory can be retained depends on the order of accuracy in the LBM
models used.  Indeed, it has been shown that certain key physical
effects beyond the Navier-Stokes equations can be accurately captured
using higher order LBM models~\cite{Shan06,Raoyang}.

There have been extensive studies in LBM for more than a
decade. However, popularly known LBM models are only accurate in the
Navier-Stokes hydrodynamic regime (c.f., \cite{fhp1,Qian}). That is,
physics higher than the Navier-Stokes order is contaminated by
numerical artifacts in these LBM models. Furthermore, there has not
been progress in systematically deriving higher order accurate LBM
models until recently~\cite{Shan06}. Originated from the framework of
the so called Lattice Gas Automata~\cite{fhp,wolf}, the conventional
approach to formulating LBM models is based on a so called ``top
down'' procedure. That is, giving a macroscopic equation such as the
Euler or the Navier-Stokes equation, an LBM model may be constructed
via an inverse Chapman-Enskog process and a {\em post-priori}
parameter matching along with various subsequent ``corrections'' (cf.,
\cite{CCM,Qian,Alexander,molvig}). But more fundamentally, because
such an approach relies on the availability of macroscopic
descriptions, it encounters an intrinsic difficulty in extending
physics beyond the original macroscopic equations. It is well known
that there is no well established and reliable macroscopic equation
for deeper non-equilibrium physics beyond the Navier-Stokes regime.

One can theoretically show that the level of non-equilibrium physics
is directly associated to the hydrodynamic
moments~\cite{Shan06}. Specifically, from the representation of the
Chapman-Enskog expansion, there exists an apparent hierarchical
relationship among hydrodynamic moments at various non-equilibrium
levels. That is, $n$-th order hydrodynamic moments at $m$-th
non-equilibrium level are related to the $(n+1)$-th order moments at
$(m-1)$-th non-equilibrium level. Carrying out this hierarchy all the
way, we see that in order to ensure the $n$-th moment physics at the
$m$-th non-equilibrium level, it requires the equilibrium moments of
$(m+n)$-th order to be accurate.  In other words, the higher order
equilibrium hydrodynamic moments captured accurately, the wider range
of non-equilibrium physics can be described. Indeed, the popularly
known lattice Boltzmann models are only accurate up to the second
order equilibrium moment (i.e., the equilibrium momentum flux
tensor). As a result, these models only give an approximately correct
``level-1'' non-equilibrium momentum flux. This is a reason why the
conventional LBM models are only applicable to the Navier-Stokes
(Newtonian) fluid physics in low Mach number isothermal
situations~\cite{wolf,fhp1,CCM,Qian}.

Based on the above, we see that the essential requirement for
accurately capturing a wider range of physics is directly related to
achieving equilibrium hydrodynamic moments to higher orders. Once the
higher order moments are accurately realized, the resulting
hydrodynamic equations such as the Euler, the Navier-Stokes and beyond
are automatically attained.  This is accomplished without the
conventional {\em post-priori} procedure. As shown in this paper, the
above requirement dictates a set of fundamental conditions on the
supporting lattice velocity basis in LBM.  That is, given an $N$-th
order moment accuracy requirement, the set of fundamental conditions
automatically defines the choice of a discrete lattice velocity set
and its corresponding weights for such a purpose.

In this paper, we theoretically derive this set of fundamental
conditions for LBM models of $N$-th order. We prove how the correct
hydrodynamic moments up to the corresponding order are realized once
the conditions are satisfied.

\section{Achieving Correct Hydrodynamic Moments via Discrete Velocities} 

According to the standard continuum Boltzmann kinetic theory, an
$n$-th order equilibrium hydrodynamic moment tensor in $D$-dimension
is defined as
\begin{equation}
  \bM^{(n)} (\bx, t)
  \equiv\int d^D \bc\underbrace{\bc\bc\cdots\bc}_n f^{eq}(\bx, \bc, t)
  \label{vmoment}
\end{equation}
Equivalently, it can be expressed in a Cartesian component form as follows,
\begin{equation}
  M^{(n)}_{i_1, i_2, \cdots , i_n} (\bx, t)
  \equiv \int d^D \bc c_{i_1} c_{i_2} \cdots c_{i_n} 
  f^{eq}(\bx, \bc, t) 
  \label{moment}
\end{equation}
where subscripts $i_1, i_2, \ldots , i_n$ are Cartesian component
indices. $c_i$ is the $i$-th Cartesian component of the microscopic
particle velocity $\bc$.  The equilibrium distribution has the
standard Maxwell-Boltzmann form,
\begin{eqnarray}
  \lefteqn{f^{eq} (\bx, \bc, t) =
  \frac {\rho (\bx, t)} {[2\pi \theta (\bx, t)]^{D/2}}}\nonumber\\
  &\times&\exp\left[- \frac {(\bc - \bu (\bx, t))^2}
    {2\theta (\bx, t)} \right]
  \label{equil}
\end{eqnarray}
where the macroscopic density, fluid velocity, and temperature are
defined, respectively
\begin{eqnarray}
  \rho (\bx, t) &=& \int d^D \bc\; f^{eq} (\bx, \bc, t) \nonumber \\
  \rho \bu (\bx, t) &=& 
  \int d^D \bc\; \bc f^{eq} (\bx, \bc, t) \nonumber \\
  D\rho \theta (\bx, t) &=& 
  \int d^D \bc\; (\bc - \bu (\bx, t))^2 f^{eq} (\bx, \bc, t)
\label{consv}
\end{eqnarray}
Apparently, the above three relations correspond to the zero-th,
first, and the trace of the second order hydrodynamic moments.  It is
well known that these three moments correspond to conservation laws
and are invariant under any local collisions.

Notice the density $\rho$ is an overall multiplier on all moments,
without loss of generality for the subsequent analysis, we set it to
unity.

Now let us define an analogous hydrodynamic moment expression in terms
of summations over discrete velocity values below,
\begin{equation}
  {\tilde \bM}^{(n)} (\bx, t)
  \equiv \sum_{\alpha = 0}^b \underbrace{
    \bc_\alpha \bc_\alpha \cdots \bc_\alpha }_n f_\alpha^{eq} (\bx, t)
  \label{vmoment-d}
\end{equation}
Or equivalently, in a Cartesian component form
\begin{equation}
  {\tilde M}^{(n)}_{i_1,i_2\cdots ,i_n} (\bx, t)
  \equiv \sum_{\alpha = 0}^b c_{\alpha , i_1} c_{\alpha , i_2} 
  \cdots c_{\alpha , i_n} f_\alpha^{eq} (\bx, t)
  \label{moment-d}
\end{equation}
In the above, we have assumed there are $b + 1$ number of discrete
$D$-dimensional vector values in the basis discrete velocity set: $\{
\bc_\alpha: \alpha = 0, \ldots, b \}$.  Similarly, we define an
analogous equilibrium distribution function,
\begin{equation}
  f_\alpha^{eq} (\bx, t) = {\bar w}_\alpha (\theta (\bx, t)) 
  \exp\left[ - \frac {(\bc_\alpha - \bu (\bx, t))^2} {2\theta (\bx, t)}\right]
  \label{dequil}
\end{equation}
where the macroscopic density, fluid velocity, and temperature are now
defined in terms of moment summations instead,
\begin{eqnarray}
  1 &=& \sum_{\alpha =0}^b f_\alpha^{eq} (\bx, t) \nonumber \\
  \bu (\bx, t) &=& 
  \sum_{\alpha =0}^b \bc_\alpha f_\alpha^{eq} (\bx, t) \nonumber \\
  D\theta (\bx, t) &=& 
  \sum_{\alpha =0}^b (\bc_\alpha-\bu)^2 f_\alpha^{eq} (\bx, t)
  \label{dconsv}
\end{eqnarray}
In the above, ${\bar w}_\alpha$ is a weighting factor that is at most
dependent on $\theta (\bx, t)$. Based on this fact, we can also
re-express the discrete equilibrium distribution (\ref{dequil}) in an
alternative and simpler form:
\begin{eqnarray}
  f_\alpha^{eq} &=& {\bar w}_\alpha(\theta) 
  \exp\left[ - \frac {(\bc_\alpha - \bu)^2} {2\theta} \right]\nonumber\\
  &=& w_\alpha (\theta) \exp\left[\frac {\bc_\alpha\cdot\bu}{\theta}\right]
  \exp\left[ - \frac {\bu^2} {2\theta}\right]
  \label{dequil-1}
\end{eqnarray}
by defining $w_\alpha (\theta ) \equiv {\bar w}_\alpha (\theta ) \exp[
  - \frac {\bc^2_\alpha } {2\theta} ]$.  Therefore, the discrete
moment definition (\ref{moment-d}) can be re-expressed as,
\begin{eqnarray}
  {\tilde M}^{(n)}_{i_1,i_2, \cdots ,i_n} &\equiv& \sum_{\alpha =0}^b
  c_{\alpha, i_1} c_{\alpha, i_2}\cdots c_{\alpha, i_n}w_\alpha(\theta)
  \nonumber\\
  &\times&\exp\left[\frac {\bc_\alpha \cdot \bu}{\theta}\right]
  \exp\left[ - \frac {\bu^2} {2\theta}\right]
  \label{moment-d1}
\end{eqnarray}

Having all the basic definitions above specified, we are now ready to
prove several fundamental conditions for a lattice velocity basis
supporting an $n$-th order hydrodynamic moment accuracy and its
corresponding form for the discrete equilibrium distribution
function. These conditions are set forth for measuring any given
lattice in terms of an intrinsic tensor:
\begin{equation}
  E^{(n)}_{i_1,\cdots ,i_n} \equiv \sum_{\alpha =0}^b 
  w_\alpha (\theta ) c_{\alpha , i_1} c_{\alpha , i_2} 
  \cdots c_{\alpha , i_n}  
  \label{tensor}
\end{equation} 

\begin{theorem}
  Discrete moment ${\tilde \bM}^{(n)}$ is equal to the moment
  $\bM^{(n)}$ of the continuum Boltzmann kinetic theory, if the
  supporting lattice velocity basis satisfy the following conditions:
  \begin{equation}
    E^{(n)}_{i_1,i_2,\cdots ,i_n} = \left\{
      \begin{array}{ll}
	\theta^{n/2}\Delta^{(n)}_{i_1,i_2,\cdots ,i_n}, & n = 0, 2, 4,
	\ldots\\
	0, & n = 1, 3, 5, \ldots
      \end{array}
    \right.
    \label{condition}
  \end{equation}
\end{theorem}

In the above, $\Delta^{(n)}_{i_1,i_2,\cdots ,i_n}$ is the $n$-th order
delta function defined as a summation of $n/2$ ($n = even\; integer$)
products of simple Kronecker delta functions $\delta_{i_1i_2} \cdots
\delta_{i_{n-1}i_n}$ and those from distinctive permutations of its
sub-indices~\cite{wolf,manifolds,group1,group2}.  There are $(n-1)!!$
($\equiv (n - 1) \cdot (n - 3) \ldots 3 \cdot 1$) total number of
distinctive terms in $\Delta^{(n)}_{i_1i_2\ldots i_n}$.  For instance,
$\Delta^{(2)}_{ij} \equiv \delta_{ij}$, and
\begin{eqnarray}
  \Delta^{(4)}_{ijkl} &=&
  \delta_{ij}\delta_{kl} +
  \delta_{ik}\delta_{jl} +
  \delta_{il}\delta_{jk} \nonumber \\
  \Delta^{(6)}_{ijklmn} &=&
  \delta_{ij}\Delta^{(4)}_{klmn} +
  \delta_{ik}\Delta^{(4)}_{lmnj} +
  \delta_{il}\Delta^{(4)}_{mnjk} \nonumber\\
  &+&\delta_{im}\Delta^{(4)}_{njkl} +
  \delta_{in}\Delta^{(4)}_{jklm}
  \label{delt}
\end{eqnarray} 
Obviously, a lattice velocity set that satisfy condition
(\ref{condition}) for $E^{(n)}$ is $n$-th order isotropic.

\vspace {.1in}

\noindent {\bf Proof of Theorem 1}: First we prove for the zero-th
order moment, ${\tilde \bM}^{(0)} = \bM^{(0)} = 1$.  According to
(\ref{dequil-1}) we have,
\begin{eqnarray}
  {\tilde \bM}^{(0)} &=& \exp\left[-\frac{\bu^2}{2\theta}\right]
  \sum_{\alpha =0}^bw_\alpha\exp\left[\frac{\bc_\alpha\cdot\bu}\theta
    \right] \nonumber \\
  &=& \exp\left[ -\frac{\bu^2}{2\theta}\right]
  \sum^\infty_{l=0} \frac {1}{\theta^l l!} \sum_{\alpha =0}^b w_\alpha 
  (\bc_\alpha \cdot \bu)^l 
  \label{m0}
\end{eqnarray}
If (\ref{condition}) is satisfied, then all odd valued $l$ terms
vanish, and the even valued terms become,
\begin{eqnarray}
  \sum_{\alpha =0}^b w_\alpha (\bc_\alpha \cdot \bu)^{2l} 
  &=& \theta^l \Delta^{(2l)} \otimes \underbrace{\bu \bu \cdots \bu}_{2l}
  \nonumber\\
  &=& (2l - 1)!! \theta^l \bu^{2l}
\end{eqnarray}
In the above $\otimes$ denotes a scalar product of two tensors.
Therefore, (\ref{m0}) reduces to,
\begin{eqnarray}
  {\tilde \bM}^{(0)} &=& \exp\left[ - \frac{\bu^2}{2\theta}\right]
  \left\{ 1 + \sum^\infty_{l = 1} \frac {(2l - 1)!!} {(2l)!}
  \frac {\bu^{2l}} {\theta^l} \right\} \nonumber \\
  &=& \exp\left[ - \frac {\bu^2} {2\theta}\right]
  \left\{ 1 + \sum^\infty_{l = 1} \frac {1} {l!}
  \frac{\bu^{2l}}{(2\theta )^l}\right\}
  \label{m0-1}
\end{eqnarray}
where the identity $ (2l-1)!!/(2l)! = 2^{-l}/l!$ is used.  Since,
\begin{equation}
  1 + \sum^\infty_{l = 1} \frac {1} {l!} \frac {\bu^{2l}} {(2\theta )^l} 
  = \exp\left[ \frac {\bu^2} {2\theta}\right]
\end{equation}
Substituting this into (\ref{m0-1}), we have proved that ${\tilde
\bM}^{(0)} = 1$.

Next, we prove ${\tilde \bM}^{(n)} = \bM^{(n)}$ for $n > 0$. We start
this by defining a partition function in discrete velocity space,
\begin{equation}
  {\cal Q} \equiv \sum_{\alpha = 0}^b w_\alpha 
  \exp\left[\frac {\bc_\alpha \cdot \bu} {\theta}\right]
  = \exp\left[ \frac {\bu^2} {2\theta}\right]
  \label{partition}
\end{equation}
Notice the second equality in the above is a result of the analysis of
${\tilde \bM}^{(0)} = 1$. Consequently, we show that satisfying the
second equality is a sufficient condition for achieving the correct
hydrodynamic moment for any integer $n$.  First of all, we have the
following general relationship,
\begin{eqnarray}
  {\tilde \bM}^{(n)} &=& \expo{- \frac {\bu^2} {2\theta}}
  \sum_{\alpha =0}^b w_\alpha 
  \underbrace{\bc_\alpha\bc_\alpha\cdots\bc_\alpha}_n 
  \exp\left[\frac{\bc_\alpha\cdot\bu}{\theta}\right] \nonumber \\
  &=& \expo{- \frac {\bu^2} {2\theta}}
  \theta^n \frac {\partial^n} {\partial \bu^n}
  \sum_{\alpha = 0}^b w_\alpha 
  \expo{\frac {\bc_\alpha\cdot\bu}\theta}
  \nonumber \\
  &=& \expo{- \frac {\bu^2} {2\theta}}
  \theta^n \frac {\partial^n} {\partial \bu^n} {\cal Q}
  \label{m-n}
\end{eqnarray}
Since ${\cal Q} = \expo{\bu^2 / 2\theta}$, then Eq.~(\ref{m-n}) becomes
\begin{equation}
  \tilde{\bM}^{(n)} = \expo{-\frac{\bu^2}{2\theta}}
  \theta^n \frac {\partial^n}{\partial \bu^n}
  \left[ \exp\left(\frac {\bu^2} {2\theta}\right) \right]
  \label{d-momt}
\end{equation}
In comparison, from the continuum Boltzmann kinetic theory, we have
\begin{eqnarray}
  \bM^{(n)} &=& \frac 1{(2\pi\theta)^{D/2}}
  \int d^D \bc\; \underbrace{ \bc \cdots \bc}_n
  \expo{-\frac {(\bc - \bu)^2} {2\theta}} \nonumber \\
  &=& e^{- \frac {\bu^2} {2\theta} } \int d^D \bc\;
  \underbrace{ \bc \cdots \bc}_n 
  (2\pi \theta )^{-\frac {D} {2}} e^{- \frac {\bc^2} {2\theta}
    + \frac {\bc \cdot \bu} {\theta}} \nonumber \\
  &=& e^{ - \frac {\bu^2} {2\theta} }
  \theta^n \frac {\partial^n} {\partial \bu^n}
  \int d^D \bc\;
  (2\pi \theta )^{-\frac {D} {2}} e^{- \frac {\bc^2} {2\theta}
    + \frac {\bc \cdot \bu} {\theta}}
  \label{c-momt}
\end{eqnarray}
It is easily shown that
\[
\int d^D \bc\;
(2\pi \theta )^{-\frac {D} {2}} e^{- \frac {\bc^2} {2\theta}
+\frac {\bc \cdot \bu} {\theta}} = \expo{\frac {\bu^2} {2\theta}}
\]
Henceforth, we have shown that (\ref{c-momt}) and (\ref{d-momt}) have
exactly the same form. Subsequently, we have proved the theorem that
${\tilde \bM}^{(n)} = \bM^{(n)}$ for any positive integer $n$, if
condition (\ref{condition}) is satisfied.

It is revealing to check a few obvious representative examples. First
of all, the first moment
\[
  {\tilde \bM}^{(1)} = \expo{-\frac{\bu^2}{2\theta}}
  \theta \frac {\partial} {\partial \bu}
  \expo{\frac{\bu^2}{2\theta}} = \bu
\]
This is simply the fluid momentum or the fluid velocity.

On the other hand, the second moment
\[
  {\tilde \bM}^{(2)} = \expo{- \frac {\bu^2} {2\theta}}
  \theta^2 \frac {\partial^2} {\partial \bu^2}
  \expo{\frac{\bu^2}{2\theta}} = \theta {\bm I} + \bu\bu
\]
where ${\bm I}$ is the second rank unity tensor. Hence the second
moment has precisely the same form of the correct hydrodynamic
momentum flux tensor.  Furthermore, we have
\[ \frac {1} {2} Trace( {\tilde \bM}^{(2)} ) 
= \frac {D} {2}\theta + \frac {1} {2}\bu^2 \]
which is exactly the hydrodynamic total energy.

\section{Moment Accuracy for Lattices of Finite Isotropy} 

In the previous section, we have proved that condition
(\ref{condition}) sufficiently ensures all moments defined via
summations over discrete lattice velocity values are equal to that of
the continuum Boltzmann kinetic theory.  However, such a condition is
unnecessarily too strong, because it requires the supporting lattice
basis to have an infinite isotropy (i.e., $n \rightarrow
\infty$). Obviously, no lattice velocity set containing a finite
number of discrete values is able to meet such a requirement.  Hence a
realistic goal is to find a relationship between the hydrodynamic
moments up to a given finite order and the corresponding isotropy for
the supporting lattice velocity basis.

First of all, we notice the existence of a hierarchical relationship
among the hydrodynamic moments.  Based on definition (\ref{m-n}), we
have
\begin{equation}
  \tilde{\bM}^{(n)} = \expo{-\frac{\bu^2}{2\theta}}
  \left(\theta \frac {\partial} {\partial \bu}\right)^n {\cal Q}
  \label{defm}
\end{equation}
Hence,
\begin{eqnarray*}
  \lefteqn{\tilde{\bM}^{(n+1)} = \expo{-\frac{\bu^2}{2\theta}}
  \theta \frac {\partial} {\partial \bu} 
  \left(\theta \frac {\partial} {\partial \bu}\right)^n {\cal Q}}\\
  &=& \expo{-\frac{\bu^2}{2\theta}}\theta\frac{\partial}{\partial\bu} 
  \left[e^{\frac{\bu^2}{2\theta}}e^{-\frac{\bu^2}{2\theta}}
  \left(\theta \frac{\partial} {\partial \bu}\right)^n {\cal Q}\right]\\
  &=& \expo{-\frac{\bu^2}{2\theta}}\theta\frac{\partial}{\partial\bu} 
  \left[\expoa{\frac{\bu^2}{2\theta}}\tilde{\bM}^{(n)}\right]
\end{eqnarray*}
This gives the hierarchical relationship,
\begin{equation}
  {\tilde \bM}^{(n+1)} = \bu {\tilde \bM}^{(n)} + 
  \theta \frac {\partial} {\partial \bu} {\tilde \bM}^{(n)} 
  \label{hiera}
\end{equation}
Using the hierarchical relationship (\ref{hiera}), all higher order
moments are derivable starting from ${\tilde \bM}^{(0)} = 1$.  More
importantly, we realize that $n$-th order moment ${\tilde \bM}^{(n)}$
is an $n$-th order polynomial in terms of the power of the fluid
velocity.  That is, the highest power in ${\tilde \bM}^{(n)}$ is
$\bu^n$. Since hydrodynamic moments up to a finite order only involve
a finite power of fluid velocity, we expect moment accuracy up to a
finite order can be achieved by a finite lattice set of adequate
isotropy.  Having established these properties, we arrive at the next
theorem below.

\vspace {.1in}

\begin{theorem}
  If the supporting lattice velocity basis satisfies the following conditions
  \begin{equation}
    E^{(n)}_{i_1,\cdots ,i_n} = \left\{
    \begin{array}{ll}
      \theta^{n/2}\Delta^{(n)}_{i_1,\cdots ,i_n},
      & n = 0, 2,\ldots, 2N\\
      0, & n = \mbox{odd integer}
    \end{array}\right.
    \label{conditionN}
  \end{equation}
  and if the discrete equilibrium distribution function $f^{eq,
    (N)}_\alpha$ is a truncation of the original exponential form by
    retaining terms only up to $\bu^N$, then the discrete moment
    ${\tilde \bM}^{(n)}$ is accurate and equal to the moment ${\bm
    M}^{(n)}$ of the continuum Boltzmann kinetic theory for any $n
    \leq N$. $N$ is any given finite positive integer.
\end{theorem} 

It is easily recognized that the basis lattice velocity set satisfying
the above condition must be $2N$-order isotropic ({\it c.f.},
\cite{wolf,orszag}).

\vspace {.1in}

\noindent {\bf Proof of Theorem 2}: We start by first examining the
standard Maxwell-Boltzmann distribution (\ref{equil}), and express it
in an expanded form in powers of fluid velocity $\bu$.  This is very
easily accomplished by taking advantage of the following generating
function for Hermite series,
\begin{equation}
  \expo{2tx - t^2} = \sum_{n = 0}^\infty \frac {H_n(x)} {n!} t^n
  \label{hermi}
\end{equation}
where $H_n(x)$ is the standard $n$-th order Hermite polynomial.  Let
us define the unity vector ${\hat u} \equiv \bu / |\bu|$ and $|\bu|
\equiv \sqrt{\sum_{i = 1}^D u^2_i}$ is the magnitude, and $\xi \equiv
\bc\cdot {\hat u}/\sqrt{2\theta}$. We can formally express the
distribution (\ref{equil}) as,
\begin{eqnarray}
  f^{eq}(\bx, \bc, t) &=& \frac 1{(2\pi \theta)^{\frac D2}}
  \expo{-\frac{(\bc - \bu)^2} {2\theta}}\nonumber \\
  &=& \frac {1} {(2\pi\theta)^{\frac {D} {2}}} e^{- \frac {\bc^2}
  {2\theta}}
  \expo{\frac {\bc \cdot \bu}\theta - \frac{\bu^2}{2\theta}}
  \nonumber \\
  &=& \frac 1{(2\pi \theta )^{\frac {D} {2}}} e^{- \frac {\bc^2} {2\theta} } 
  \sum_{n = 0}^\infty\frac {H_n (\xi )} {n!} 
  \left(\frac \bu {\sqrt{2\theta}}\right)^n
\end{eqnarray}
A truncated series $f^{eq, (N)}$ of the above can be defined by simply
retaining the terms up to $\bu^N$.  Based on the orthogonal property
of the Hermite polynomials, namely
\begin{equation}
  \int^\infty_{-\infty} dx \; e^{-x^2} H_m(x) H_n(x) = 0; \quad
  \forall m \neq n
\end{equation}
and because $\bc$ of power $N$ can be fully represented by Hermite
polynomials $\{ H_n ; \; n = 0, \ldots , N \}$, it is straightforward
to see that moments up to $N$-th order constructed out of $f^{eq}$ are
identical to that of $f^{eq, (N)}$, for the higher order terms in
$f^{eq}$ give vanishing contributions due to orthogonality.

Next, similar to the above, we expand the discrete distribution (\ref{dequil-1}), and keeping terms only up to $\bu^N$,
\begin{eqnarray}
  f_\alpha^{eq, (N)}
  &=& w_\alpha\expo{\frac {\bc_\alpha \cdot \bu} {\theta}  
    - \frac {\bu^2} {2\theta}} \nonumber \\
  &=&  w_\alpha \sum_{n = 0}^N 
  \frac {H_n (\xi_\alpha)}{n!}\left(\frac\bu{\sqrt{2\theta}}\right)^n
  \label{inf-expand}
\end{eqnarray}
where $\xi_\alpha \equiv \bc_\alpha \cdot {\hat u}
/\sqrt{2\theta}$. Hence the task to prove Theorem 2 is to prove
${\tilde \bM}^{(n)}$ ($\forall n \leq N$) generated by
$f_\alpha^{eq,(N)}$ is equal to $\bM^{(n)}$ from the full
Maxwell-Boltzmann distribution $f^{eq}$ or its truncation $f^{eq,
(N)}$.  According to definition (\ref{vmoment}), we have
\begin{eqnarray}
  \bM^{(n)} &\equiv& \int d^D \bc\; \underbrace{\bc \cdots \bc}_n
  f^{eq}(\bx, \bc, t)\nonumber \\
  &=& \frac 1{(2\pi \theta )^{D/2}}\sum_{n=0}^N\frac 1{n!}
  \left(\frac\bu{\sqrt{2\theta}}\right)^n\nonumber \\ 
  &\times& \int d^D \bc\; \underbrace{\bc \cdots \bc}_n
  \expo{-\frac{\bc^2}{2\theta}}H_n(\xi)
  \label{Nmoment}
\end{eqnarray}
On the other hand, according to (\ref{vmoment-d}), we have
\begin{eqnarray}
  \lefteqn{\tilde{\bM}^{(n)} \equiv \sum_{i=0}^b 
  \underbrace{\bc_\alpha \cdots \bc_\alpha}_n f_\alpha^{eq, (N)}}
  \nonumber\\
  &=& \sum_{n = 0}^N\frac 1{n!}\left( \frac\bu{\sqrt{2\theta}}\right)^n
  \sum_{\alpha =0}^b\underbrace{\bc_\alpha \cdots \bc_\alpha }_n w_\alpha 
  H_n (\xi_\alpha)
  \label{Nmoment-d}
\end{eqnarray}
>From (\ref{Nmoment}) and (\ref{Nmoment-d}), we see that both of these
involve Hermite polynomials of orders no greater than
$N$. Furthermore, a given Hermite function $H_n(x)$ is a polynomial of
$x^m$ ($m = 0, \ldots , \leq n$). Therefore, both $\bM^{(n)}$ and
${\tilde \bM}^{(n)}$ involve powers of $\bc$ (or $\bc_\alpha$) from 0
up to $n + N$.  Based this observation, we see that it is sufficient
to prove ${\tilde \bM}^{(n)} = \bM^{(n)}$ ($\forall n \leq N$), if for
all integer $m \leq 2N$ the following property is satisfied,
\begin{equation}
\int d^D \bc \; \frac {exp[- {\bc^2} /2\theta ]} {(2\pi \theta )^{D/2}} 
\underbrace{\bc \cdots \bc}_m 
= \sum_{\alpha = 0}^b w_\alpha
\underbrace{\bc_\alpha \cdots \bc_\alpha}_m
\label{ced}
\end{equation}
\[ \forall m = 0, \ldots , 2N \]
The result for the discrete summation is already given in the
definition of (\ref{tensor}) and (\ref{conditionN}).  Hence it is
suffice to just show that this is also true for the continuum
integration.  In fact, according to the basic Gaussian integral
property, we know that
\begin{eqnarray}
  \lefteqn{\frac 1{(2\pi\theta)^{D/2}}
    \int d^D\bc\expo{-\frac{\bc^2}{2\theta}}
 c_{i_1} c_{i_2} \cdots c_{i_m}} \nonumber \\
  &=& \left\{
  \begin{array}{ll}
    \theta^{m/2} \Delta^{(m)}_{i_1,i_2,\cdots ,i_m}, 
    & m = 0, 2, 4, \ldots , 2N\\
    0,&  m = 1, 3, 5, \ldots , 2N+1
  \end{array}
  \right.
  \label{quadra}
\end{eqnarray}
Consequently, we have proved ${\tilde \bM}^{(n)} = \bM^{(n)}$
($\forall n \leq N$), and thus Theorem 2.

It is also worthwhile to note, without repeating the explicit steps of
the above, that the same proof applies if the truncation of the
exponential form $f^{eq}_\alpha$ is up to $N + 1$. Thus, we can retain
an extra term in the expanded form.

\section{Discussion}

In this paper, we have presented and proved a set of fundamental
conditions for formulating LBM models.  Lattice velocity sets obeying
these conditions automatically produce equilibrium moment accuracy to
any given $N$-th order.  As demonstrated in~\cite{Shan06},
non-equilibrium moments are theoretically expressible as spatial and
temporal derivatives of equilibrium moments. Therefore, achieving
higher order moment accuracy enables accurate description of fluid
properties into deeper non-equilibrium regimes~\cite{Chapman,jfm}.
This is essential for physical properties at finite Knudsen or Mach
numbers that are beyond the Navier-Stokes representation.

To make a more direct comparison with conventional LBM models, we
rewrite (\ref{inf-expand}) in a more explicit form (up to $O(u^5)$)
below,
\begin{eqnarray}
  f_\alpha^{eq} &=& w_\alpha \rho \; 
  [1 + \frac {\bc_\alpha \cdot \bu} {\theta}  
    + \frac {(\bc_\alpha \cdot \bu)^2} {2\theta^2} 
    - \frac {\bu^2} {2\theta} 
    \nonumber \\ 
    & & + \frac {(\bc_\alpha \cdot \bu)^3} {6\theta^3} 
    - \frac {(\bc_\alpha \cdot \bu)\bu^2} {2\theta^2} 
    \nonumber \\
    & & + \frac {(\bc_\alpha \cdot \bu)^4} {24\theta^4} 
    - \frac {(\bc_\alpha \cdot \bu)^2\bu^2} {4\theta^3} 
    + \frac {\bu^4} {8\theta^2} 
    \nonumber \\ 
    & & + \frac {(\bc_\alpha \cdot \bu)^5} {120\theta^5} 
    - \frac {(\bc_\alpha \cdot \bu)^3\bu^2} {12\theta^4} 
    + \frac {(\bc_\alpha \cdot \bu)\bu^4} {8\theta^3} ] 
  \label{mach}
\end{eqnarray} 
It is immediately recognized that the series for most of the
conventional LBM models terminate at $O(u^2)$ or $O(u^3)$.  For
example, the so called D3Q15 and D3Q19 correspond to the expansion up
to $O(u^2)$~\cite{Qian}. It can be directly verified that their
underlying lattice velocity sets only satisfy the fundamental
conditions (\ref{conditionN}) up to $N = 2$, so that the higher order
moment terms beyond $O(u^3)$ can not be accurately
supported. Furthermore, in these models, the temperature is fixed at
$\theta = 1/3$.  An extended 34-velocity model
exists~\cite{molvig,ChenTeixeira}, and its temperature has a range of
variation between $1/3$ to $2/3$, and D3Q19 is its reduced limit as
$\theta = 1/3$. But the moment accuracy is still $N = 2$.

There are typically two approaches to construct lattice velocity sets
obeying higher order of accuracies ($N > 2$) according to
(\ref{conditionN}).  One approach is to rely on relations between
discrete rotational symmetry and tensor
isotropy~\cite{wolf,orszag}. For instance, we can start with a lattice
velocity set consisting of multiple lattice speeds, namely
\begin{equation}
  {\cal L} = {\cal L}_1 \cup {\cal L}_2 \cdots \cup {\cal L}_M
\end{equation}
where each of the subset is defined as
\[ 
{\cal L}_\beta = \{ {\bm c}_{\alpha , \beta} ; \; i = 0, \ldots , b_\beta \} 
\]
\[ \beta = 1, \ldots , M \]
All lattice velocities in each subset ${\cal L}_\beta$ has the same
magnitude, $|{\bm c}_{\alpha , \beta}| = c_\beta$. This way, the
required isotropy can be imposed at each speed level.  It has been
shown that if such a velocity subset is parity invariant and obeys an
$n$-th order isotropy ($n = even \; integer$), then its basic moment
tensor has the following form~\cite{orszag}
\begin{equation}
  {\bm E}^{(n),\beta}_{i_1, i_2, \cdots , i_n} = 
  b_\beta c^n_\beta \frac {(D - 2)!!} {(D + n - 2)!!} 
  \Delta^{(n)}_{i_1, i_2, \cdots , i_n}
\end{equation}
and it vanishes for all the odd integer moments. Subsequently, we can
assign a weighting factor $w_\beta (\theta)$ for each subset ${\cal
L}_\beta$, so that the overall condition (\ref{conditionN}) is
achieved by satisfying the following constraint on the weighting
factors,
\begin{equation}
  \sum_{\beta = 1}^M b_\beta c^n_\beta \frac{(D - 2)!!} {(D + n - 2)!!} 
  w_\beta(\theta) = \theta^{n/2}
\end{equation} 
for $n = 0, 2, \ldots , 2N$. There are $2N + 1$ such
constraints. Hence, it is necessary to include enough number of
subsets and $w_\beta (\theta )$ ($\beta = 1, \ldots , M \geq N+1$) in
order to have a solution. Using such a procedure, a 59-velocity model
in 3-dimension is formulated that satisfies (\ref{conditionN}) up to
$N = 3$ with 6-th order tensor isotropy, so that the expansion in
(\ref{mach}) can be carried to $O(u^4)$. Based on the analysis above
and else where~\cite{Shan06}, such an order of moment accuracy is
necessary for getting the correct energy flux in thermal
hydrodynamics~\cite{yuchen,teixeira,wateri}. Another approach is to
form the discrete velocity sets via Gaussian quadrature for higher
order models~\cite{Shan06}.  Indeed, (\ref{quadra}) defines the
precise requirement. The only difference here is that the quadratures
need to allow a variable temperature $\theta$. This approach is
relatively more straightforward, so that it enables a systematic
formulation of higher accurate LBM models to 6-th, 8-th orders and
beyond. There is also a similar work recently by Sbragaglia et al on
how to construct higher order isotropic moments~\cite{Sbrag}.

The formulation described in this paper offers a rigorous measure for
evaluating the order of accuracy of a given LBM model.  For future
convenience, we may simply refer an LBM model that satisfies condition
(\ref{conditionN}) to $N$-th order as ``E(N)-accurate.''

\vspace{0.1in}

\noindent {\bf Acknowledgments}: We dedicate this work to the
celebration of 250 years of Euler Equation. We are grateful to Steven
Orszag and Raoyang Zhang for their valuable discussions. This work is
supported in part by the National Science Foundation.

\end{document}